\documentclass[showpacs,10pt,twocolumn,prb]{revtex4-1}
\usepackage{amsmath}
\usepackage{amssymb}
\usepackage{graphics}
\usepackage{epsfig}
\usepackage{CJK}
\usepackage{color}
\usepackage{lineno}

\begin{document}


\title{Enhanced superconductivity and anisotropy of FeTe$_{0.6}$Se$_{0.4}$ single crystals with Li-NH$_{3}$ intercalation}
\author{Chenghe Li$^{1,\dag}$, Shanshan Sun$^{1,\dag}$, Shaohua Wang$^{1}$, and Hechang Lei$^{1,*}$}
\affiliation{$^{1}$Department of Physics and Beijing Key Laboratory of Opto-electronic Functional Materials $\&$ Micro-nano Devices, Renmin University of China, Beijing 100872, China}
\date{\today}

\begin{abstract}
We report a systematic study of anisotropy resistivity, magnetoresistance and Hall effect of Li$_{0.32}$(NH$_{3}$)$_{y}$Fe$_{2}$Te$_{1.2}$Se$_{0.8}$ single crystals. When compared to the parent compound FeTe$_{0.6}$Se$_{0.4}$, the Li-NH$_{3}$ intercalation not only increases the superconducting transition temperature, but also enhances the electronic anisotropy in both normal and superconducting states. Moreover, in contrast to the parent compound, the Hall coefficient $R_{\rm H}$ becomes negative at low temperature, indicating electron-type carriers are dominant due to Li doping. On the other hand, the sign reverse of $R_{\rm H}$ at high temperature and the failure of scaling behavior of magnetoresistance imply that hole pockets may be still crossing or just below the Fermi energy level, leading to the multiband behavior in Li$_{0.32}$(NH$_{3}$)$_{y}$Fe$_{2}$Te$_{1.2}$Se$_{0.8}$.
\end{abstract}

\maketitle

\section{Introduction}

The study of iron-based superconductors (IBSCs) is one of hotspots in the field of condensed matter physics and material science. Among the family of IBSCs, $\beta-$FeSe has attracted much attention because of its simple crystal structure, which is preferable for comprehending the superconducting mechanism of IBSCs, as well as unique properties such as the presence of superconductivity ($T_{c}=$ 8 K) without carrier doping \cite{Hsu} and dramatic pressure effect on $T_{c}$ \cite{Medvedev}. For FeSe-based SCs, the isovalent substitution is an effective way to tune physical properties at superconducting and normal states. For example, substituting Se with Te in FeSe can increase $T_{c}$ to about 15 K \cite{Yeh}. With substituting Se with S, the structural (nematic) transition at $\sim$ 87 K is suppressed gradually and the $T_{c}$ is slightly enhanced up to 10 K\cite{Watson,Abdel-Hafiez}. These results suggest that the isovalent substitution can change electronic structure subtly through introducing chemical pressure and/or bonding ionicity/covalency, as in FeAs-based SCs \cite{Wang}. Beside isovalent substitution, carrier doping via intercalation of alkali/alkaline earth/rare earth metals and (Li, Fe)OH layers in between FeSe layers is another important way to take effects on superconducting properties. $T_{c}$ can be greatly enhanced to about 30 - 45 K with electron doping \cite{Guo,Lu,Ying}. For these heavily electron doped FeSe-based SCs, there are only electron pockets near the Brillouin zone corners ($M$ point) \cite{Zhao2}, distinctly different from $\beta-$FeSe where both hole pockets around the $\Gamma$ point and electron pockets around the $M$ point exist \cite{Nakayama}. In contrast, the results of theoretical calculations and angle-resolved photoemission spectroscopy indicate that the isovalent substitution does not change the topology of Fermi surface significantly \cite{Watson,Subedi,Chen}. Thus, whether both series of FeSe-based SCs have common pairing mechanism and symmetry is still an open question.

Because both of isovalent substitution and carrier doping can increase $T_{c}$ to varying extents, it would be interest to study the evolution of superconductivity when applying both methods simultaneously. Previous studies on K$_{x}$Fe$_{2-y}$Se$_{2-z}$S$_{z}$, Na$_{0.80}$(NH$_{3}$)$_{0.6}$Fe$_{1.86}$(Se$_{1-z}$S$_{z}$)$_{2}$, and (Li$_{0.8}$Fe$_{0.2}$)OHFeSe$_{1-x}$S$_{x}$ indicate that although the $T_{c}$ decreases with S substitution \cite{Lei,Guo2,Lu2}, they are still much higher than those in FeSe$_{1-x}$S$_{x}$. On the other hand, the study of carrier doping effects on FeSe$_{1-x}$Te$_{x}$ is scarce. For example, superconductivity is suppressed quickly when substituting Se with Te in Rb$_{0.8}$Fe$_{2-y}$Se$_{2-x}$Te$_{x}$ and the $T_{c}$ disappears completely when $x=$ 0.4 \cite{Gu}. Another example is (Li/Na)$_{x}$(NH$_{3}$)$_{y}$Fe$_{2-\delta}$(Se$_{1-z}$Te$_{z}$)$_{2}$ that also exhibits the suppression of superconductivity with increasing the content of Te. However, the $T_{c}$ is about 21 K at $z=$ 0.5, higher than FeTe$_{0.5}$Se$_{0.5}$ ($T_{c}\sim$ 15 K) \cite{Lei2,Yeh}. Due to the powder form of samples, the detailed physical properties, especially transport properties, are still unknown.

Recently, we have grown Li$_{x}$(NH$_{3}$)$_{y}$Fe$_{2}$Se$_{2}$ (LiFeSe-122) single crystals successfully by using the low-temperature ammonothermal method \cite{Sun}. In this work, we grown Te substituted Li$_{0.32}$(NH$_{3}$)$_{y}$Fe$_{2}$Te$_{1.2}$Se$_{0.8}$ (LiFeTeSe-122) single crystals based on this method and report a detailed study on their transport properties. We find that the electron-type carriers are dominant at low temperature, confirming the electron doping from Li to Fe(Te, Se) layers. Moreover, superconductivity and anisotropy of physical properties of LiFeTeSe-122 are greatly enhanced when compared to the parent compound FeTe$_{0.6}$Se$_{0.4}$, but smaller than those in LiFeSe-122.

\section{Experiment}

The LiFeTeSe-122 single crystals were synthesized by the low-temperature ammonothermal technique \cite{Sun,WSH}. X-ray diffraction (XRD) patterns were collected using a Bruker D8 x-ray Diffractometer with Cu $K_{\alpha}$ radiation ($\lambda=$ 0.15418 nm) at room temperature. The elemental analysis was performed using the inductively coupled plasma atomic emission spectroscopy (ICP-AES). Magnetization measurements were performed in a Quantum Design magnetic property measurement system (MPMS-S3). Electrical transport measurements were carried out in a Quantum Design physical property measurement system (PPMS-14). The longitudinal and Hall electrical resistivity were measured using a four-probe method on rectangular shaped single crystals. The current flowed in the $ab$ plane of crystal. The Hall resistivity was obtained from the difference of the transverse resistivity measured at the positive and negative fields in order to remove the longitudinal resistivity contribution due to voltage probe misalignment, i.e., $\rho_{xy}(\mu_{0}H)=[\rho(+\mu_{0}H)-\rho(-\mu_{0}H)]/2$. The $c$-axis resistivity $\rho_{c}(T)$ was measured by attaching current and voltage wires on the opposite sides of the plate-like crystal.

\section{Results and Discussion}

\begin{figure}[tbp]
\centerline{\includegraphics[scale=0.3]{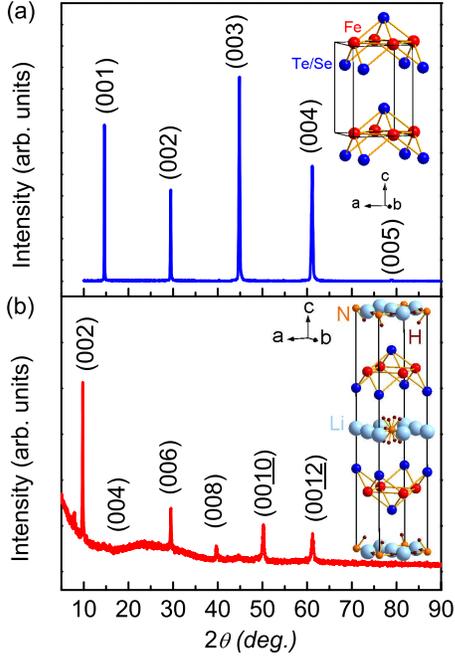}} \vspace*{-0.3cm}
\caption{XRD patterns and crystal structures of (a) FeTe$_{0.6}$Se$_{0.4}$ and (b) LiFeTeSe-122 single crystals.}
\end{figure}

The crystal structure of FeTe$_{0.6}$Se$_{0.4}$ is illustrated in the inset of Fig. 1(a). FeTe$_{0.6}$Se$_{0.4}$ has a tetragonal structure with P4/nmm space group (No. 129). The key structural units are Fe(Te, Se) layers, where Fe ions form the simple square lattice and chalcogen ions (Se and Te) are placed in the centers of these squares, above and below the Fe plane in chess-board order. When intercalating Li-NH$_{3}$ in between Fe(Te, Se), the space group of LiFeTeSe-122 becomes I4/mmm (No. 139), isostructural to LiFeSe-122, and in each unit cell, there are two layers of Fe(Te, Se) (inset of Fig. 1(b))\cite{Burrard-Lucas,Lei2}. For both single crystals, only (00l) reflections can be indexed (Fig. 1(a) and (b)), indicating that the surfaces of crystals are parallel to the $ab$ plane. Compared to FeTe$_{0.6}$Se$_{0.4}$, the diffraction peaks of LiFeTeSe-122 shift to lower angle because of the expansion of interlayer distance of Fe(Te, Se) layers after intercalation. Moreover, the refinement of the powder XRD pattern for LiFeTeSe-122 indicates that the $a$-axial lattice parameter also becomes slightly larger than FeTe$_{0.6}$Se$_{0.4}$, i.e., the Fe(Te, Se) layer is stretched after intercalation\cite{Lei2,WSH}.
The atomic ratio of Li : Fe : Te : Se determined from the ICP-AES analysis is 0.16 : 1.00 : 0.60 : 0.38. Based on this result, the estimated electron doping level is $\sim$ 0.16 e/Fe.

\begin{figure}[tbp]
\centerline{\includegraphics[scale=0.24]{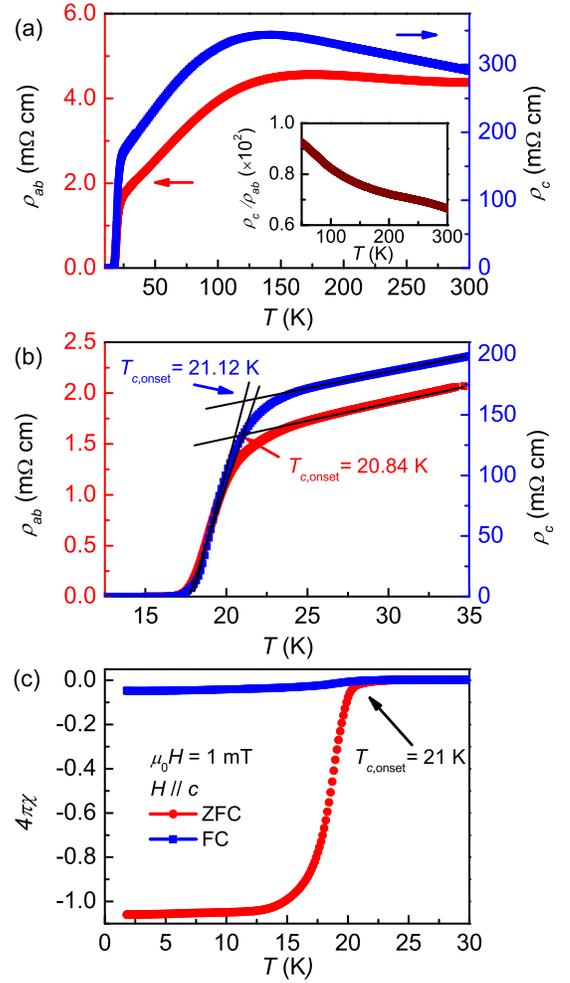}} \vspace*{-0.3cm}
\caption{(a) Temperature dependence of in-plane resistivity $\rho_{ab}(T)$ and out-of-plane resistivity $\rho_{c}(T)$ at zero field. Inset: the ratio of $\rho_{c}/\rho_{ab}$ as a function of temperature. (b) The enlarged part of $\rho_{ab}(T)$ and $\rho_{c}(T)$ at low temperature region. (c) Temperature dependence of dc magnetic susceptibility 4$\pi\chi(T)$ up to 30 K with zero-field-cooling and field-cooling modes ($\mu_{0}H=$ 1 mT, $H \Vert c$).}
\end{figure}

Fig. 2(a) shows the temperature dependence of in-plane resistivity $\rho_{ab}(T)$ and out-of-plane resistivity $\rho_{c}(T)$ for LiFeTeSe-122 single crystal at zero field. The $\rho_{ab}(T)$ is almost temperature-independent at $T>$ 150 K and then exhibits a metallic behavior below 150 K. In contrast, the temperature dependence of $\rho_{c}(T)$ is non-monotonic, i.e., a semiconductor-like behavior when $T>$ 125 K and a metallic behavior at $T<$ 125 K. These behaviors of $\rho_{ab}(T)$ and $\rho_{c}(T)$ are very similar to those in FeTe$_{0.6}$Se$_{0.4}$ single crystals \cite{Noji}. But the anisotropy of resistivity $\rho_{c}$/$\rho_{ab}$ is larger than that in the latter. It is about 66 at the 300 K and increases to about 92 with decreasing temperature to 50 K, comparing with about 70 at 50 K for FeTe$_{0.6}$Se$_{0.4}$ \cite{Noji}. It indicates an enhanced anisotropy of resistivity at normal state in LiFeTeSe-122 single crystals due to the intercalation of Li-NH$_{3}$. On the other hand, the ratio of $\rho_{c}$/$\rho_{ab}$ for LiFeTeSe-122 is smaller than that of Li$_{x}$(NH$_{3}$)$_{y}$Fe$_{2}$Se$_{2}$ ($\sim$ 8000 at 50 K)\cite{Sun}, suggesting that Te doping increases the interlayer interaction and weakens the two-dimensionality of samples. As shown in Fig. 2(b), there are sharp resistivity drops appearing in the $\rho_{ab}(T)$ and $\rho_{c}(T)$ curves at zero filed and they correspond to the superconducting transitions. The $T_{c}$ for $\rho_{ab}(T)$ and $\rho_{c}(T)$ is 20.84 K and 21.22 K with the transition width $\Delta T_{c}=$ 2.97 K and 3.02 K, respectively. Temperature dependence of zero-field-cooling magnetic susceptibility $4\pi\chi(T)$ at $\mu_{0}H=$ 1 mT along the $c$-axis (Fig. 2(c)) further confirms the sharp superconducting transition with $T_{c,\rm onset}=$ 21 K. Besides, the much smaller superconducting volume fraction derived from field-cooling $4\pi\chi(T)$ curve at 2 K suggests a strong vortex pinning in the sample.
\begin{figure}[tbp]
\centerline{\includegraphics[scale=0.16]{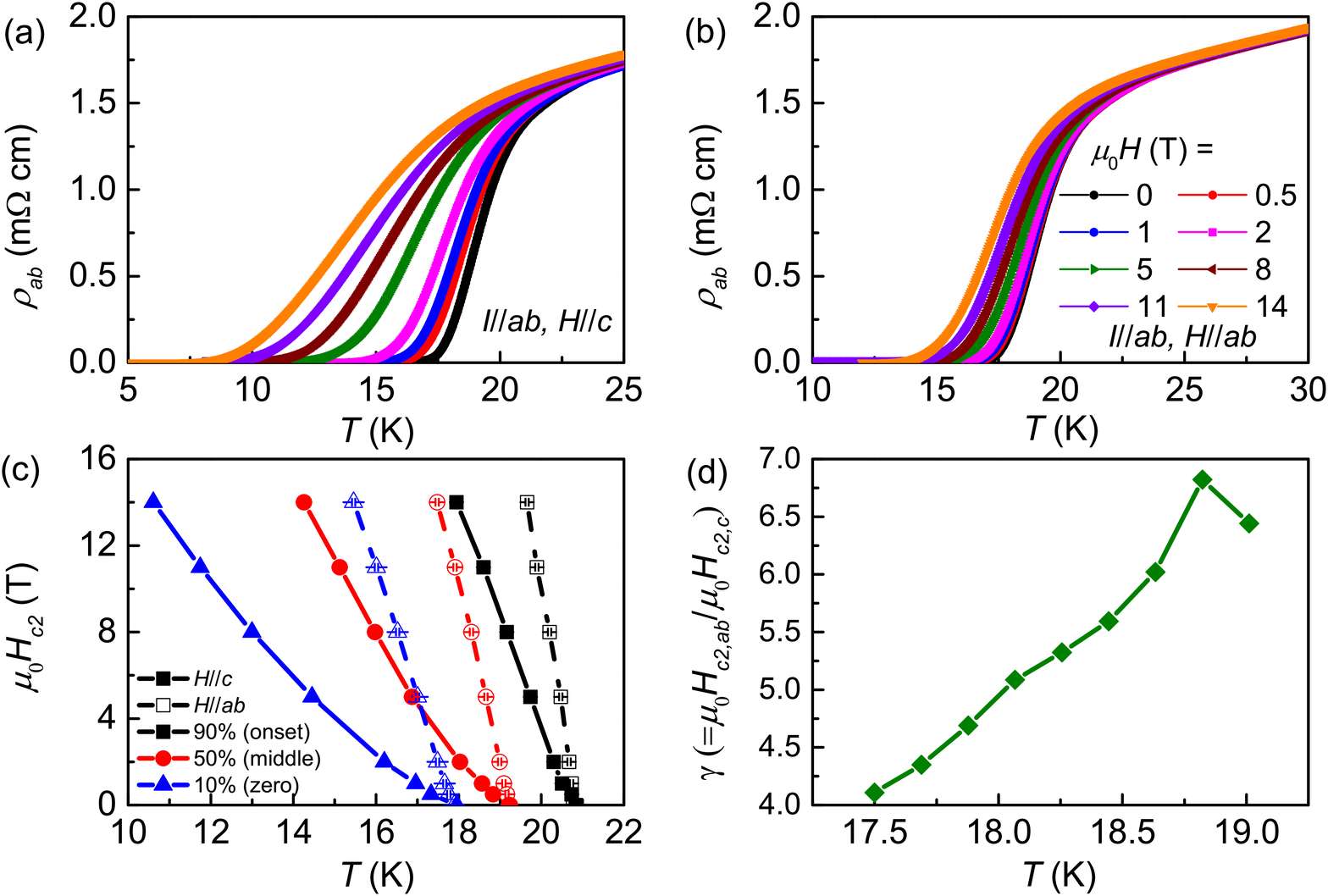}} \vspace*{-0.3cm}
\caption{Temperature dependence of $\rho _{ab}(T)$ at various magnetic fields for (a) $H\Vert ab$ and (b) $H\Vert c$. (c) Temperature dependence of resistive upper critical fields $\mu_{0}H_{c2}(T)$ corresponding to three criterions for both field directions. (d) Temperature dependence of the anisotropy of upper critical field $\gamma=\mu_{0}H_{c2,ab}/\mu_{0}H_{c2,c}$ using the 50\% $\rho_{n}$ criterion.}
\end{figure}

The temperature dependence of $\rho _{ab}(T)$ of LiFeTeSe-122 single crystal at different magnetic fields up to 14 T for $H\Vert c$ and $H\Vert ab$ are shown in Fig. 3(a) and (b), respectively. With increasing fields, the transition width becomes broader at higher fields, especially for $H\Vert c$, implying a strong flux-flow behavior as observed in NdFeAsO$_{0.7}$F$_{0.3}$ and high-temperature $T_{c}$ cuprates\cite{Jaroszynski,Lee,Fendrich}. The temperature dependence of upper critical field $\mu_{0}H_{c2}(T)$ determined using the 90 \%, 50 \% and 10 \% drop of $\rho_{n}$ (normal state $\rho_{ab}$ at the transition temperature) are shown in Fig. 3(c).
All curves of $\mu_{0}H_{c2}(T)$ are almost linear in temperature, except that of $\mu_{0}H_{c2,\rm zero}(T)$ for $H\Vert c$ which exhibits an obvious upturn behavior near $T_{c,\rm onset}(0)$. The slopes of $\mu_{0}H_{c2}(T)$ at $T_{c,\rm onset}(0)$, $T_{c,\rm middle}(0)$ and $T_{c,\rm zero}(0)$ are -11.5, -9.0, and -5.9 T/K for $H\Vert ab$ and -5.1, -3.2, and -2.1 T/K for $H\Vert c$, respectively. When $H\Vert ab$, the slope of $\mu_{0}H_{c2,\rm middle}(T)$ near $T_{c,\rm middle}(0)$ for LiFeTeSe-122 is larger than that of FeTe$_{0.6}$Se$_{0.4}$ (-7.2 T/K), but when $H\Vert c$, it is smaller than that in the latter (-4.9 T/K) \cite{Lei3}.
The zero-temperature values of $\mu_{0}H_{c2}(0)$ can be estimated by using the Werthamer-Helfand-Hohenberg (WHH) formula\cite{WHH}, $\mu_{0}H_{c2}(0)=-0.693 T_{c}(d\mu_{0}H_{c2}/dT)\mid_{T_{c}}$. We estimate the $\mu_{0}H_{c2,\rm middle}(0)$ is 119.8 T and 42.8 T for $H\Vert ab$ and $H\Vert c$, respectively. When compared to LiFeSe-122, The Te doping decreases the $\mu_{0}H_{c2,\rm middle}(0)$ for $H\Vert ab$ but increases that for $H\Vert c$ \cite{Sun}.
The anisotropy of $\mu_{0}H_{c2}$ defined as $\gamma=\mu_{0}H_{c2,ab}/\mu_{0}H_{c2,c}$ is shown in Fig. 3(d). The values of $\gamma$ displays a notable decrease when $T$ is away from $T_{c}$(0), changing from about 6.5 at $T=$ 19 K to about 4 at $T=$ 17.5 K.

\begin{figure}[tbp]
\centerline{\includegraphics[scale=0.16]{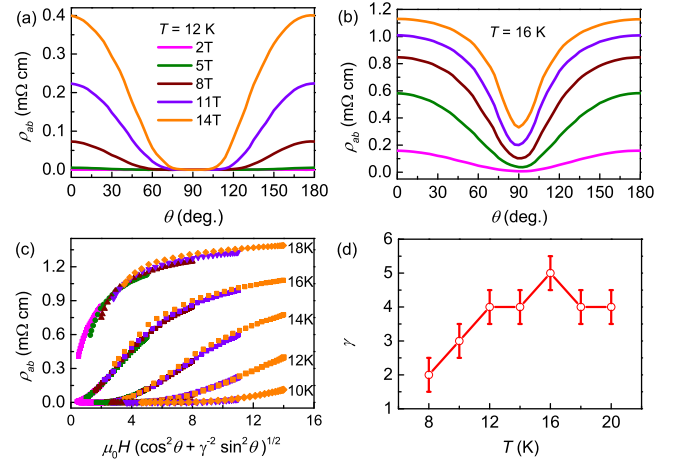}} \vspace*{-0.3cm}
\caption{Angular dependence of $\rho_{ab}(\theta, \mu _{0}H)$ at (a) 12 K and (b) 16 K with $\mu _{0}H$ = 2, 5, 8, 11, and 14 T for LiFeTeSe-122 single crystals. (b) Scaling behavior of $\rho_{ab}(\theta, \mu _{0}H)$ versus $\mu_{0}H_{s}=\mu_{0}H(\cos^{2}\theta +\gamma ^{2}\sin^{2}\theta)^{1/2}$ from 10 K to 18 K at different magnetic fields. (c) Temperature dependence of anisotropy factor $\gamma$.}
\end{figure}

The angular dependent resistivity $\rho_{ab}(\theta, \mu _{0}H)$ at various fields at $T=$ 12 and 16 K for the LiFeTeSe-122 single crystal are presented in Fig. 4(a) and (b). All of resistivity curves show a symmetric cup-like shape with the maximum values locating at $\theta=$ 0$^{\circ}$ and 180$^{\circ}$ ($\theta$ is the angle between the field direction and $c$ axis). It suggests the smaller $\mu_{0}H_{c2,c}$ than $\mu_{0}H_{c2,ab}$. According to the anisotropic Ginzburg-Landau theory based on the effective mass model \cite{Blatter,Morris}, the $\mu_{0}H_{c2}^{GL}$ is defined by $\mu_{0}H_{c2}^{GL}(\theta)=\mu_{0}H_{c2,ab}/(\sin^{2}\theta+\gamma^{2}\cos^{2}\theta)^{1/2}$. The $\rho_{ab}(\theta, \mu _{0}H)$ at different magnetic fields can be scaled to one curve through adjusting the anisotropy parameter $\gamma$. For LiFeTeSe-122, this scaling behavior is clearly observed for all of $\rho_{ab}(T,\mu _{0}H_{s})$ curves, where $\mu_{0}H_{s}=\mu_{0}H/(\sin^{2}\theta+\gamma^{2}\cos^{2}\theta)^{1/2}$ (Fig. 4(c)). The temperature dependence of $\gamma(T)$ deduced by this method is shown in Fig. 4(d). It decreases with decreasing temperature and the values are close to those determined from the ratio of $\mu_{0}H_{c2,ab}/\mu_{0}H_{c2,c}$.
The values of $\gamma(T)$ are significantly smaller than those in LiFeSe-122 ($\sim$ 8 - 16) \cite{Sun}, but still much larger than the values of $\gamma$ in FeTe$_{0.6}$Se$_{0.4}$ ($\sim$ 1 - 2) \cite{Lei3}. It suggests that although the Te doping weakens the two-dimensionality of electronic structure for Li$_{x}$(NH$_{3}$)$_{y}$Fe$_{2}$(Te$_{z}$Se$_{1-z}$)$_{2}$, the Li-NH$_{3}$ cointercalation still leads to rather large anisotropy when compared to the parent compounds FeTe$_{0.6}$Se$_{0.4}$.

\begin{figure}[tbp]
\centerline{\includegraphics[scale=0.24]{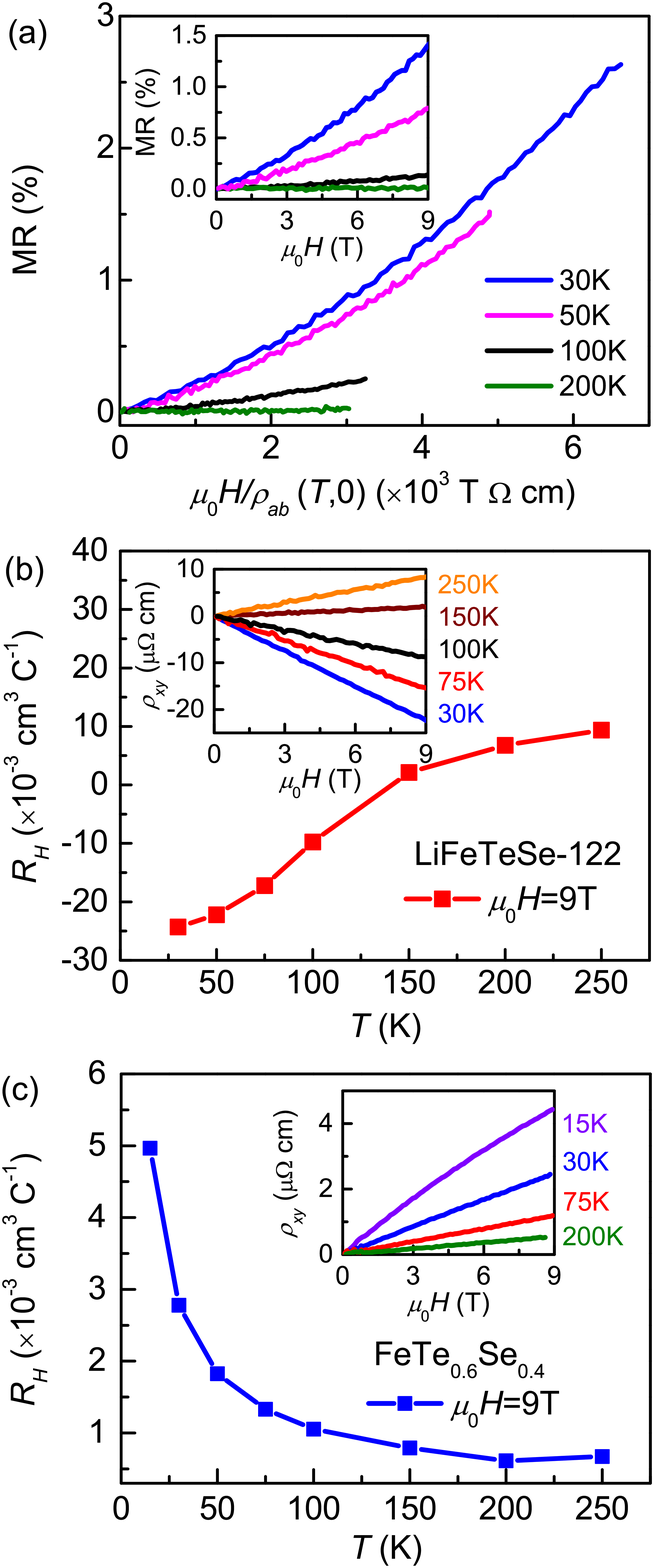}} \vspace*{-0.3cm}
\caption{(a) Kohler plot of MR between 30 - 200 K for LiFeTeSe-122 single crystals. Inset: field dependence of MR at different temperatures. Temperature dependence of the Hall coefficient $R_{\rm H}(T)$ at $\mu_{0}H$ = 9 T of (b) LiFeTeSe-122 and (c) FeTe$_{0.6}$Se$_{0.4}$ single crystals. Insets of (b) and (c): field dependence of Hall resistivity $\rho_{xy}(\mu_{0}H)$ at various temperatures.}
\end{figure}

Next, we investigate the transport properties at normal state. Field dependence of magnetoresistance (MR) [$=(\rho_{ab}(T,\mu _{0}H)-\rho_{ab}(T,0))/\rho_{ab}(T,0)$] at various temperatures are shown in the inset of Fig. 5(a). The MR is rather weak at low temperature and the magnitude is only about 1.5 \% at 30 K and 9 T, even weaker than that in LiFeSe-122 \cite{Sun}. It decreases gradually with increasing temperature. Moreover, as shown in the main panel of Fig. 5(a), the MR of LiFeTeSe-122 does not follow the Kohler's law MR = $f(\mu_{0}H\tau)=F(\mu_{0}H/\rho_{ab}(T,0))$, which will be held if there is an isotropic relaxation time $\tau$ at all points on the FS in a single-band system \cite{Pippard}. The violation of Kohler's law in LiFeTeSe-122 single crystal suggests that there might be multiple electron and hole pockets with anisotropic $\tau$ in this system, as shown in the Hall measurements.

In the whole temperature region, Hall resistivity $\rho_{xy}(\mu_{0}H)$ of LiFeTeSe-122 single crystals shows rather good linear relation against magnetic field up to 9 T (inset of Fig. 5(b)). The derived Hall coefficients $R_{\rm H}=\rho_{xy}/\mu_{0}H$ at 9 T exhibits strong temperature dependence (Fig. 5(b)). The $R_{\rm H}$ is negative below 125 K and the absolute values decreases continuously with increasing temperature. Finally the $R_{\rm H}$ becomes positive at higher temperature, i.e., there is a sign change at $T\sim$ 125 K.
It suggests the existence of two different types of charge carriers in LiFeTeSe-122 and the dominant carriers are electron-type at the low temperature and become hole-type at high temperature. In order to figure out the influence of Li-NH$_{3}$ intercalation on its electronic structure, the field dependences of $\rho_{xy}(\mu_{0}H)$ at various temperatures for parent FeTe$_{0.6}$Se$_{0.4}$ single crystals are also measured (inset of Fig. 5(c)). Similar to LiFeTeSe-122, there is a nearly linear relationship between $\rho_{xy}$ and field. The $R_{H}$ at 9 T decreases with increasing temperature but it is always positive (Fig. 5(c)), indicating the hole-type carriers dominate in FeTe$_{0.6}$Se$_{0.4}$ in the whole temperature range. The different sign of $R_{\rm H}$ at low temperature for LiFeTeSe-122 and FeTe$_{0.6}$Se$_{0.4}$ clearly shows that Li-NH$_{3}$ intercalation transfers electron from Li to Fe(Te, Se) layers. Moreover, it has to be mentioned that for undoped FeSe, the sign of $R_{\rm H}$ is negative at low temperature \cite{Watson2}. On the other hand, the $R_{\rm H}$ of LiFeTeSe-122 at $T$ just above $T_{c}$ (30 K) is much larger than that in LiFeSe-122 (about 4.8$\times$10$^{-3}$ cm$^{3}$ C$^{-1}$ at 50 K) \cite{Sun}. It indicates that the apparent carrier concentration $n_{H}$ ($=1/eR_{\rm H}=n_{h}-n_{e}$, where $n_{h}$ and $n_{e}$ are the carrier concentrations of hole and electron pockets.) in LiFeTeSe-122 (2.8$\times$10$^{20}$ cm$^{-3}$ at 30 K) is much smaller than that in LiFeSe-122 (1.3$\times$10$^{21}$ cm$^{-3}$ at 50 K). Above results suggest that in parent compound FeTe$_{0.6}$Se$_{0.4}$, Te substitution may increase the sizes of hole pockets and decrease the mobilities of electron pockets, leading to the dominance of hole-type carriers. When intercalating Li-NH$_{3}$, the doped electrons increase the sizes of electron pockets larger than hole pockets, but the latter may be still crossing or just below the Fermi energy level ($E_{F}$). It results in the significant small values of $n_{H}$ and sign reverse of $R_{\rm H}$ at high temperature when compared to LiFeSe-122 \cite{Sun}. Moreover, the violation of Kohler's law and strong temperature dependence of $R_{\rm H}$ could be partially ascribed to the multiband effect in LiFeTeSe-122.




\section{Conclusion}

In summary, we study the transport properties of Li-NH$_{3}$ intercalated FeTe$_{0.6}$Se$_{0.4}$ single crystals in detail. The $T_{c}$ is enhanced from 15 K to 21 K with carrier doping. For LiFeTeSe-122, both anisotropies of resistivity at normal state and upper critical field at superconducting state become remarkably larger than those in the parent compound. It suggests that the Li-NH$_{3}$ intercalation leads to the enhancement of anisotropy of electronic structure. In contrast to LiFeSe-122, the anisotropy of transport properties decreases, implying that Te substitution has an opposite effect when compared to carrier doping. Because the electron transfer from Li to Fe(Te, Se) layers increases the sizes of electron pockets, it results in the negative $R_{\rm H}$ at low temperature when compared to the positive one in FeTe$_{0.6}$Se$_{0.4}$. Moreover, the violation of Kohler's law and the strong temperature dependence of $R_{\rm H}$ with sign change at high temperature could be partially ascribed to the multiband effect, due to the existence of hole pockets near $E_{F}$.

\section{Acknowledgements}

This work was supported by the Ministry of Science and Technology of China (2016YFA0300504), the National Natural Science Foundation of China (No. 11574394, 11774423), the Fundamental Research Funds for the Central Universities, and the Research Funds of Renmin University of China (RUC) (15XNLF06, 15XNLQ07).

$\dag$ These authors contributed equally to this work.

$\ast$ hlei@ruc.edu.cn.

\end{document}